\documentclass[twocolumn,prl,showpacs]{revtex4}

\usepackage{epsfig}
\usepackage{graphics}
\usepackage{graphicx}
\usepackage{dcolumn}
\usepackage{amsmath,amsfonts,amssymb,amsxtra} 
\usepackage{mathrsfs}
\usepackage{bm}
\usepackage{times}

\def\MgB2{MgB$_{2}$}
\def\NbSe2{NbSe$_{2}$}
\def\cm-1{cm$^{-1}$\,}
\def\cmT-1{cm$^{-1}$/T\,}
\def\E2g{$E_{2g}$}
\def\A1g{$A_{1g}$}
\def\2DS{$2\Delta_{S}^{E}$}

\def\2DA{$2\Delta^{A}$}
\def\D0{$2\Delta_{0}$}

\begin{document} 

\title{Observation of Leggett's collective mode in a multi-band 
\MgB2 superconductor} 

\author{G.~Blumberg$^{1,*}$}
\author{A. Mialitsin$^{1}$}
\author{B.\,S. Dennis$^{1}$}
\author{M.\,V. Klein$^{2}$}
\author{N.\,D.~Zhigadlo$^{3}$}
\author{J.~Karpinski$^{3}$}

\affiliation{
$^{1}$Bell Laboratories, Alcatel-Lucent, Murray Hill, NJ 07974 \\
$^{2}$Department of Physics and Materials Research  
Laboratory,~University~of~Illinois~at~Urbana-Champaign,~Urbana,~IL~61801 \\
$^{3}$Solid~State~Physics~Laboratory,~ETH,~CH-8093~Z\"urich,~Switzerland}

\date{March 30, 2007}

\begin{abstract}

    We report observation of Leggett's collective mode in a  
multi-band \MgB2 superconductor with T$_c$\,=\,39\,K arising from the 
fluctuations in the relative phase between two superconducting 
condensates.
The novel mode is observed by Raman spectroscopy at 9.4\,meV in the 
fully symmetric scattering channel. 
The observed mode frequency is consistent with theoretical 
considerations based on the first principle computations.

\end{abstract}

\pacs{74.70.Ad, 74.25.Ha, 74.25.Gz, 78.30.Er}

\maketitle

The problem of collective modes in superconductors is almost
as old as the microscopic theory of superconductivity.
Bogolyubov \cite{Bogolyubov} and Anderson \cite{Anderson} first 
discovered that density
oscillations can couple to oscillations of the phase of
the superconducting (SC) order parameter (OP) \textit{via} the 
pairing interaction.
In a neutral system these are the Goldstone sound-like oscillations 
which accompany the spontaneous gauge-symmetry breaking, 
however,  
for a charged system the frequency of these modes is
pushed up to the plasma frequency by the Anderson-Higgs mechanism 
\cite{Anderson63} and the Goldstone mode does not exist.
The collective oscillations of the amplitude of the SC OP  
have a gap, which was first observed by
Raman spectroscopy in \NbSe2 \cite{Klein,Littlewood-Varma},  
and which plays a role equivalent to
the Higgs particle in the electro-weak theory \cite{Higgs}.  
Several other collective 
excitations have been proposed, including 
an unusual one that corresponds to fluctuations of the 
relative phase of coupled SC condensates first 
predicted by Leggett \cite{Leggett}. 
The Leggett mode is a longitudinal excitation resulting from equal 
and opposite displacements of the two superfluids along the direction 
of the mode's wavevector $q$.  
In the ideal case considered by Leggett, the mode is ``massive" and 
its energy (mass) at $q=0$ is below twice the smaller of the two gap 
energies.  
In this Letter we report the observation of Leggett's collective mode in the  
multi-band  superconductor \MgB2 with T$_c$\,=\,39\,K \cite{Akimitsu}.
The novel mode is observed in Raman response at 9.4\,meV, consistent 
with the theoretical evaluations.

\begin{figure}[b]
\begin{center}
\includegraphics[width=0.65\columnwidth]{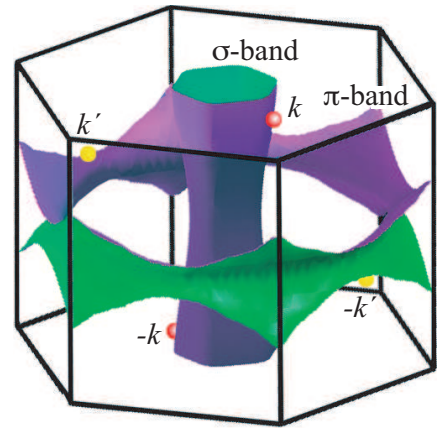}
\end{center}
\vspace{-5mm}
\caption{(color online)
An illustration of the \MgB2 Fermi surface in the first Brillouin 
zone adapted from Ref. \cite{FS}.  
A nearly cylindrical sheet of the FS around the $\Gamma - A$ line 
results from the $\sigma$-band. 
The $\pi$-band forms a FS of planar honeycomb tubular networks. 
For clarity only a single FS for each $\sigma$- and $\pi$-band pair is 
shown \cite{Liu}. 
In the SC state the $\sigma$-band Cooper pairs are bound stronger 
than the $\pi$-band pairs, at the binding energies $2\Delta_{\sigma}$ 
and $2\Delta_{\pi}$ correspondingly. 
Leggett's collective mode originates from dynamic scattering
of the $\sigma$-band pairs of electrons (illustrated in red) with 
momentum ($k, -k$) into the $\pi$-band electron pairs (yellow) with 
momentum ($k^{\prime}, -k^{\prime}$). 
} 
\label{fig:1a}
\end{figure}

The multi-gap nature of superconductivity in \MgB2 was first 
theoretically predicted \cite{Liu} and has been
experimentally established by a number of spectroscopies. 
A double-gap structure in the quasi-particle energy spectra was 
determined from tunneling spectroscopy \cite{Sza30,Iavarone}. 
The two gaps have been assigned by means of ARPES \cite{Tsuda,Souma} 
to distinct Fermi surface (FS) sheets (Fig.\,\ref{fig:1a}) belonging 
to distinct quasi-2D $\sigma$-bonding states of the 
boron $p_{x,y}$ orbitals and 3D $\pi$-states of the 
boron $p_z$ orbitals:   
$\Delta_{\sigma} =  5.5 - 6.5$ and $\Delta_{\pi} = 1.5 - 2.2$\,meV. 
Scanning tunneling microscopy (STM) has provided a reliable fit for 
the smaller gap, $\Delta_{\pi} = 2.2\,{\rm meV}$ \cite{Esklidsen}. 
This value manifests in the absorption threshold
energy at 3.8\,meV obtained from
magneto-optical far-IR studies \cite{Perucchi}.
The larger $2\Delta_{\sigma}$ gap has been demonstrated by Raman 
experiments as a SC coherence peak at about 13\,meV \cite{Quilty}.  

\begin{figure*}[t]
\begin{center}
\includegraphics[width=1.8\columnwidth]{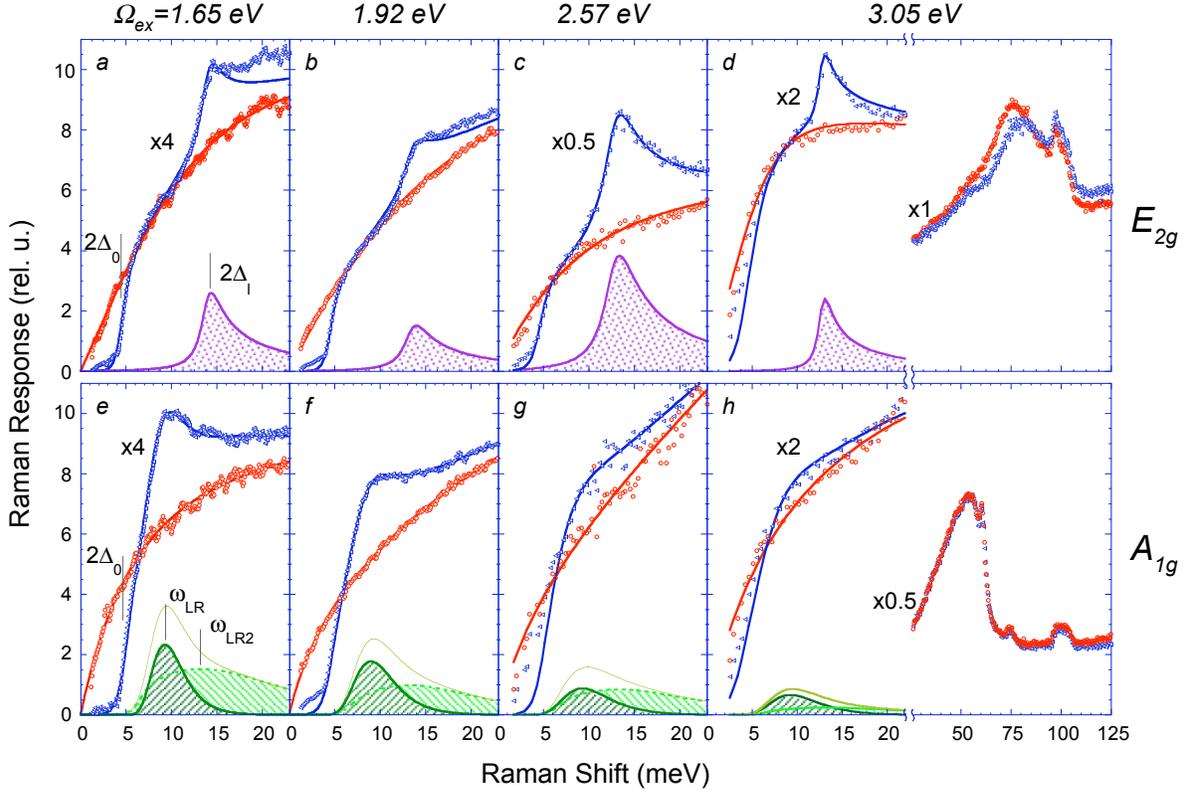}
\end{center}
\vspace{-5mm}
\caption{(color online)
The Raman response spectra of an \MgB2 crystal in 
the normal (red) and SC (blue) states for the $E_{2g}$ (top row) and 
$A_{1g}$ (bottom row) scattering channels. 
The $E_{2g}$ scattering channel is accessed by $RL$ (\emph{a}-\emph{c}) or 
$V\!H$ (\emph{d}) scattering polarization geometries and the $A_{1g}$ 
channel by $RR$ (\emph{e}-\emph{h}) geometry. 
The low temperature data is acquired at 5 - 8\,K. 
The normal state has been achieved either by increasing the crystal
temperature to 40~K (\emph{d}) or by applying a 5~T
magnetic field parallel to the $c$-axis
(\emph{a}-\emph{c},\,\emph{e}-\emph{h}) \cite{LT24}. 
The columns are arranged in the order of increasing 
excitation energy $\Omega_{ex}$. 
Solid lines are fits to the data points. 
The normal state continuum is fitted with 
$\omega/\sqrt{a + b \omega^{2}}$ functional form.  
The data in the SC state is decomposed into a sum of a 
gapped normal state continuum with temperature 
broadened $2\Delta_{0}= 4.6$\,meV gap cutoff,
the SC coherence peak at $2\Delta_{l} = 13.5$\,meV (shaded in violet), 
and the collective modes at $\omega_{LR} = 9.4$\,meV and $\omega_{LR2} = 
13.2$\,meV (shaded in dark and light green). 
The solid hairline above the shaded areas is the sum of both modes. 
To fit the observed shapes the theoretical BCS coherence peak singularity 
$\chi^{\prime\prime} \sim 4 \Delta_{l}^{2}/(\omega \sqrt{\omega^{2} - 
4 \Delta_{l}^{2}})$ is broadened by convolution with a Lorentzian with 
HWHM = 5 - 12\% of $2\Delta_{l}$ \cite{KleinDierker}. 
The collective mode $\omega_{LR}$ is fitted with the response 
function shown in Fig.\,\ref{fig:4}. 
Panels (\emph{d} and \emph{h}) also show the high energy part of 
spectra for respective symmetries. 
The broad $E_{2g}$ band at about 79\,meV corresponds to the boron 
stretching mode that is the only phonon that exhibits 
renormalization below the SC transition \cite{Mialitsin}. 
For the $A_{1g}$ channel the spectra are dominated by two-phonon 
scattering. 
} 
\label{fig:1}
\end{figure*}
Polarized Raman scattering measurements from the $ab$ 
surface of \MgB2 single crystals grown as described in
\cite{Karpinski} were performed in back scattering geometry using
less than 2\,mW of incident power focused to a $100 \times 200\,
\mu$m spot. The data in a magnetic field were acquired with a
continuous flow cryostat inserted into the horizontal bore of a SC
magnet. The sample temperatures quoted have been corrected for
laser heating. 
We used the excitation lines of a Kr$^{+}$ laser and a
triple-grating spectrometer for analysis of the scattered light.
The data were corrected for the spectral response of the
spectrometer and the CCD detector and for the optical properties of 
the material at different wavelengths as described in Ref.
\cite{Blumberg94}.

The factor group associated with ${\rm MgB_2}$ is $D_{6h}$.
We denote by
$(\textbf{e}_{in} \textbf{e}_{out})$ a configuration in which the
incoming/outgoing photons are polarized along the
$\textbf{e}_{in}$/$\textbf{e}_{out}$ directions. The vertical
($V$) or horizontal ($H$) directions were chosen
perpendicular or parallel to the crystallographic $a$-axis. The
''right-right'' ($RR$) and ''right-left'' ($RL$)
notations refer to circular polarizations:
$\textbf{e}_{in} = (H - i V) / \sqrt{2}$, with
$\textbf{e}_{out} = \textbf{e}_{in}$ for the $RR$ and
$\textbf{e}_{out} = \textbf{e}_{in}^{*}$ for the $RL$ geometry. 
For the $D_{6h}$ factor group the $RR$ polarization scattering 
geometry selects the $A_{1g}$ symmetry while both $RL$ and $V\!H$ select
the $E_{2g}$ representation. 

\begin{figure}[t]
\begin{center}
\includegraphics[width=0.85\columnwidth]{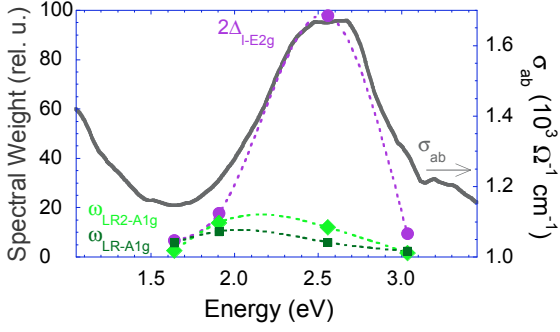}
\end{center}
\vspace{-5mm}
\caption{(color online)
The comparisons of the $ab$-plane optical 
conductivity \cite{Kuz'menko} $\sigma_{ab}$ (solid line) to the 
integrated spectral weight under SC coherence peaks as a function of 
excitation energy:  
$2\Delta_{l}$ in the $E_{2g}$ (violet circles) and 
Leggett's collective modes $\omega_{LR}$ (dark green squares) and 
$\omega_{LR2}$ (light green diamonds) in the $A_{1g}$ channel. 
All dashed lines are guides for the eye. 
} 
\label{fig:2}
\end{figure}

Light can couple to electronic and phononic excitations \emph{via} resonant or 
non-resonant Raman processes \cite{DevereauxRMP}. 
The Raman scattering cross-section can be substantially 
enhanced when the incident 
photon energy is tuned into resonance with optical interband 
transitions. 
For \MgB2 the interband contribution to the in-plane optical conductivity  
$\sigma_{ab}(\omega)$ contains strong IR peaks with a tail 
extending to the red part of the visible range and 
a pronounced resonance around 2.6~eV \cite{Kuz'menko} 
(Fig.\,\ref{fig:2}).   
The IR peaks are associated with transitions between two 
$\sigma$-bands while the peak in the visible 
range is associated with a transition from the $\sigma$ band
to the $\pi$ band \cite{Kuz'menko,Kortus}.

In Fig.\,\ref{fig:1} we show the Raman response from an \MgB2 
single crystal for the \E2g and \A1g scattering channels
for four excitation photon energies in the normal and SC states.  
Besides the phononic scattering at high Raman shifts all spectra show 
a moderately strong featureless electronic Raman continuum. 
The origin of this continuum is likely due to finite wave-vector 
effects \cite{KleinDierker,DevereauxRMP,ratio}. 
For isotropic single band metals the Raman response in the fully  
symmetric channel is  
expected to be screened \cite{KleinDierker,DevereauxRMP,AbrikosovGenkin}. 
However, for \MgB2 the electronic scattering intensity in the 
$A_{1g}$ and $E_{2g}$ channels is almost equally strong.
  
The low frequency part of the 
electronic Raman continuum changes in the SC state 
(Fig.\,\ref{fig:1}), reflecting renormalization of electronic 
excitations resulting in four new features in the spectra:  
(i) a threshold of Raman intensity at $2\Delta_{0}= 4.6$\,meV, 
(ii) a SC coherence peak at $2\Delta_{l} = 13.5$\,meV in the \E2g 
channel, and two new modes in the \A1g channel,   
(iii) at 9.4\,meV, which is in-between the $2\Delta_{0}$ and 
$2\Delta_{l}$ energies, and
(iv) a much broader mode just below $2\Delta_{l}$. 
The observed energy scales of the fundamental gap $\Delta_{0}$ 
and the large gap  $\Delta_{l}$ are consistent with 
$\Delta_{\pi}$ and $\Delta_{\sigma}$ as assigned by one-electron 
spectroscopies \cite{Tsuda,Souma,Esklidsen}. 

(i) At the fundamental gap value $2\Delta_{0}$ the spectra for both 
symmetry channels show a threshold without a coherence peak. 
This threshold is cleanest for the spectra with lower energy photon
excitations $\Omega_{ex}$ for which the low-frequency contribution of 
multi-phonon scattering from acoustic branches is suppressed \cite{Mialitsin}. 
The absence of the coherence peak above the threshold is consistent 
with the expected behavior for a superconductor with SC coherence length 
larger than the optical penetration depth \cite{KleinDierker}.

(ii) The $2\Delta_{l}$ coherence peak appears in the \E2g channel  
as a sharp singularity with  
continuum renormalization extending to high energies, which 
agrees with expected behavior for clean 
superconductors \cite{DevereauxRMP,KleinDierker,ratio}. 
The Raman coupling to this mode is provided by density-like 
fluctuations in the $\sigma$-band hence the peak intensity is 
enhanced by about an order of magnitude when the excitation photon 
energy $\Omega_{ex}$ is in resonance with the 2.6~eV  
$\sigma \rightarrow \pi$ inter-band transitions (Fig.\,\ref{fig:2}).  

(iii) The novel peak at 9.4\,meV is observed only in the \A1g 
scattering channel. 
This mode 
is more pronounced for off-resonance excitation for which the 
electronic continuum above the fundamental threshold 
$2\Delta_{0}$ is weaker. 
We assign this feature to the collective mode proposed by 
Leggett \cite{Leggett}:
If a system contains two coupled superfluid liquids a simultaneous 
cross-tunneling of a pair of electrons becomes possible 
(Fig.\,\ref{fig:1a}). 
Leggett's collective mode is caused by counter flow of the two 
superfluids leading to small fluctuations of the relative phase of 
the two condensates while the total electron density is locally 
conserved.  
In a crystalline superconductor, its symmetry is that of the fully 
symmetric irreducible representation of 
the group of the wave-vector $q$. 
If the energy of this mode is below the smaller pair-breaking gap 
energy, dissipation is suppressed and the excitation should be 
long-lived.  
In the case of \MgB2 the two coupled SC condensates reside at the 
$\sigma$- and $\pi$-bands.
The oscillation between the condensates involves the scattering
of a pair of $\sigma$-band electrons with momentum ($k, -k$) into a 
pair of $\pi$-band electrons with momentum ($k^{\prime}, -k^{\prime}$) due 
to the interaction between the electrons. 
\begin{table}[t]
\caption{
Estimates of Leggett's mode frequency $\omega_{L}$, the vertex 
correction $\omega_{V}$ and the Raman resonance frequency 
$\omega_{LR}$ based on values of intra- and inter-band pairing 
potentials $V_{ij}$ ($i,j = \sigma, \pi$) deduced from first 
principal calculations (two band model)
\cite{Liu,MazinReview,Choi,Golubov}.  
The effective density of states $N_{\sigma} =  2.04$ and $N_{\pi} =  
2.78$\,Ry$^{-1}$spin$^{-1}$cell$^{-1}$ \cite{Liu} and 
the experimental values for the SC gaps $\Delta_{\sigma} =  6.75$ and 
$\Delta_{\pi} = 2.3$\,meV are used.}
\begin{center}
\begin{tabular}{l|cccccc}
\hline
 
   & $V_{\sigma\sigma}$
   & $V_{\pi\pi}$
   & $V_{\sigma\pi}$
   & $\omega_{L}$ 
   & $\omega_{V}$ 
   & $\omega_{LR}$ \\
    Refs. & (Ry) & (Ry) & (Ry) & (meV) & (meV) & (meV) \\  
\hline\hline
Liu \emph{et al.} \cite{Liu}         & 0.47 & 0.1 & 0.08 & 6.2 & 7.1 
& 7.9\\
Choi \emph{et al.} \cite{Choi}       & 0.38 & 0.076 & 0.054 & 6.2 & 
6.7 & 7.8\\
Golubov \emph{et al.} \cite{Golubov} & 0.5 & 0.16 & 0.077 & 5.1 & 5.7 
& 6.9\\
\hline
\end{tabular}
\end{center}
\label{tab:1}
\end{table}
The Leggett mode is gapped (``massive"). 
Its dispersion for small momentum $q$ obeys relation \cite{Leggett,Sharapov} 
\begin{equation}
    \Omega_{L}(q)^{2} = \omega_{L}^{2} + v^{2}q^{2}, 
    \label{dispersion}
\end{equation}
where the excitation gap $\omega_{L}$ is given by solution of  
\cite{MVKlein}
\begin{equation}
   L(\omega)^{2} = \omega^{2}  
    \label{LeggettModeEq}
\end{equation}
with 
\begin{equation}
   L(\omega)^{2} = 
             \frac{4 \Delta_{\sigma} \Delta_{\pi} 
   V_{\sigma\pi}}{\det{V}}  
\frac{N_{\sigma} f_{\sigma}({\omega}) + N_{\pi} f_{\pi}({\omega})}
       {N_{\sigma} f_{\sigma}({\omega}) N_{\pi} 
       f_{\pi}({\omega})}.    
\label{LEq}
\end{equation}
Here $V$ is the matrix of intra- and inter-band interaction with 
pairing potentials $V_{\sigma\sigma}$, $V_{\pi\pi}$ and 
$V_{\sigma\pi}$; 
$N_{\sigma}$ and $N_{\pi}$ are the density of states in 
corresponding bands; and we define a complex function 
$f_{\sigma,\pi}(\Tilde{\omega}) = 
    \frac{\arcsin{\Tilde{\omega}}}{\Tilde{\omega} \sqrt{1 - 
    \Tilde{\omega}^{2}}}$, 
with $\Tilde{\omega} = \omega/2 \Delta_{\sigma,\pi}$.     
The solution for Leggett's mode Eq.\,(\ref{LeggettModeEq}) exists if  
\begin{equation}
\det{V} > 0. 
    \label{condition}
\end{equation}
If $\omega_{L} \ll \min({\Delta_{\sigma}, \Delta_{\pi}})$ 
it reduces to the original Leggett expression \cite{Leggett,Sharapov}
\begin{equation}
    \omega_{L}^{2} = \frac{N_{\sigma} + N_{\pi}}{N_{\sigma} 
    N_{\pi}} \frac{4 V_{\sigma\pi} \Delta_{\sigma} 
    \Delta_{\pi}}{\det{V}}. 
    \label{LeggettMode}
\end{equation}
Because it is fully symmetric with respect to symmetry operations 
that leave the wave-vector $q$ invariant this mode contributes only 
to the \A1g Raman response. 
Because its neutrality the mode remains unscreened by Coulomb 
interactions.   
Generalization of Eqs. (10a-c) and (18) from 
Ref.\,\cite{KleinDierker} to the two band case \cite{MVKlein} give 
Raman response:  
\begin{equation}
    \chi_{A_{1g}}(\omega) = 
    -\frac{8 \Delta_{\sigma} \Delta_{\pi} 
   V_{\sigma\pi}}{\det{V}} 
    \frac{(\gamma_{\sigma}-\gamma_{\pi})^{2}}
    {L(\omega)^{2} + \omega_{V}^{2} - 
    \omega^{2}}.  
    \label{chi}
\end{equation}
Here $\gamma_{\sigma,\pi}$ are the bare light coupling vertices for 
corresponding bands and 
$\omega_{V}^{2} = 
{4 \Delta_{\sigma} \Delta_{\pi} V_{\sigma\pi} (V_{\sigma\sigma} + 
V_{\pi\pi} - 2 V_{\sigma\pi})}/{\det{V}}$ 
is due to the vertex correction \cite{MVKlein}. 
For light to couple to Leggett's excitation $\gamma_{\sigma}$ and 
$\gamma_{\pi}$ should not be equal, the coupling is further enhanced 
if $\gamma_{\sigma} \gamma_{\pi} < 0$. 
The latter condition is satisfied for \MgB2 since the $\sigma$-bands 
are hole-like while the $\pi$-bands are predominantly electron-like.  
The integrated intensity of the Leggett's mode as a function of 
excitation energy does not follow the optical 
conductivity and is about five times weaker than the resonantly enhanced  
coherence peak in the \E2g channel (Fig.\,\ref{fig:2}).  

\begin{figure}[t]
\begin{center}
\includegraphics[width=0.8\columnwidth]{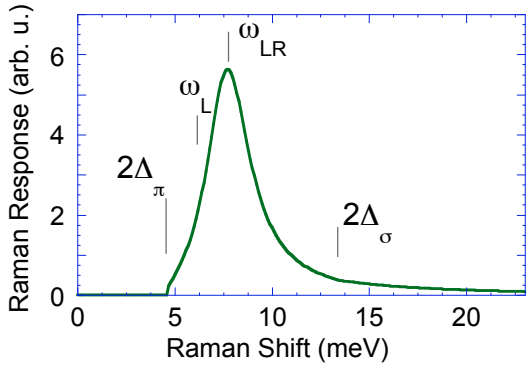}
\end{center}
\vspace{-5mm}
\caption{(color online)
Im\,$\chi_{A_{1g}}(\omega)$ given by Eq.\,(\ref{chi}) using the 
interaction matrix by Liu \emph{et al.} \cite{Liu}. 
} 
\label{fig:4}
\end{figure}
The estimates of the two-band interaction matrices by first principle 
computations \cite{Liu,MazinReview,Choi} which are collected in 
Table~\ref{tab:1} show that for \MgB2 the  
condition (\ref{condition}) is satisfied. 
In Fig.\,\ref{fig:4} we show the calculated Raman response function 
(\ref{chi}) for the first set of parameters from Table \ref{tab:1} in 
the $q \rightarrow 0$ limit. 
Finite wave-vector contribution from the $\pi$-band will stretch the 
$\pi$-band Raman continuum in agreement with the data. 
Model calculations suggest that interference with the $\sigma$-band 
coherence peak might produce a structure at about $2\Delta_{l}$. 
We note that the estimates for bare Leggett's mode frequency 
$\omega_{L}$ are close to the $\sim6.2$\,meV value observed by 
tunneling spectroscopy \cite{Brinkman}   
and the estimates for the peak in Raman response (\ref{chi}), 
$\omega_{LR}$, are consistent with the observed mode at 9.4\,meV. 
Because the collective mode energy is between the two-particle 
excitation thresholds 
for $\pi$- and $\sigma$-band, 
$2\Delta_{\pi} < \omega_{L} < \omega_{LR} < 2\Delta_{\sigma}$, 
Leggett's excitation relaxes into the $\pi$-band continuum. 
Indeed, the measured $Q$-factor for this mode is about two: 
the mode energy relaxes into the $\pi$-band quasiparticle continuum 
within a couple of oscillations.   

(iv) Finally we note that \MgB2 has four FSs, two nearly 
cylindrical sheets due to the $\sigma$-bands split and two 
tubular network structures originate from $\pi$-bands. 
Solution to the Leggett problem extended to 4-bands with $4 
\times 4$ interaction matrix given by Liu \emph{et al.} \cite{Liu} 
leads to two Raman resonances: $\omega_{LR} = 8.4$\,meV and second 
$\omega_{LR2}$ just 0.05\,meV below the $2\Delta_{l}$ gap. 
We interpret the superconductivity induced intensity in the \A1g 
channel just below the $2\Delta_{l}$ energy as evidence either for a 
second Leggett resonance or for interference between SC contributions 
from the $\pi$-band with large-$qv_{Fc}$ and the $\sigma$-band 
with small $qv_{Fc}$. 
A sum of two modes peaking at 9.4 and 13.2\,meV with very similar 
excitation profiles provides a good fit to the experimental data.

We conclude that despite being short lived, Leggett excitations in 
\MgB2 are observed in \A1g Raman response. 

The authors thank D. van der Marel, I. Mazin and W. E. Pickett for 
valuable discussions. 
AM was supported by the Lucent-Rutgers Fellowship program. 
NDZ was supported by the Swiss National Science Foundation through 
NCCR pool MaNEP.

\end{document}